# GEOGRAPHIC CONCENTRATION IN PORTUGAL AND REGIONAL SPECIFIC FACTORS

**Vítor João Pereira Domingues Martinho**

Unidade de I&D do Instituto Politécnico de Viseu
Av. Cor. José Maria Vale de Andrade
Campus Politécnico
3504 - 510 Viseu
**(PORTUGAL)**
e-mail: vdmartinho@esav.ipv.pt

**ABSTRACT**

This paper pretends to analyze the importance which the natural advantages and local resources are in the manufacturing industry location, in relation with the "spillovers" effects and industrial policies. To this, we estimate the Rybczynski equation matrix for the various manufacturing industries in Portugal, at regional level (NUTS II) and for the period 1980 to 1999. Estimations are displayed with the model mentioned and for four different periods, namely 1980 to 1985, from 1986 to 1994, from 1980 to 1994 and from 1995 to 1999. The consideration of the various periods until 1994, aims to capture the effects of our entrance at the, in that time, EEC (European Economic Community) and the consideration of a period from 1995 is because the change in methodology for compiling statistical data taken from this time in Portugal. As a summary conclusion, noted that the location of manufacturing in Portugal is still mostly explained by specific factors, with a tendency to increase in some cases the explanation by these factors, having the effect "spillovers" and industrial policies little importance in this context.

**1. INTRODUCTION**

In this work, taking into account the work of (1)Kim (1999), seeks to analyze the importance of the natural advantages and local resources (specific factors of locations) have in explaining the geographic concentration over time in the Portuguese regions, relatively effects "spillovers" and industrial policies (in particular, the modernization and innovation that have allowed manufacturing in other countries take better advantage of positive externalities). For this, we estimated the Rybczynski equation matrix for the different manufacturing industries in the regions of Portugal, for the period 1980 to 1999. It should be noted that while the model of inter-regional trade, the Heckscher-Ohlin-Vanek, presents a linear relationship between net exports and inter-regional specific factors of locations, the Rybczynski theorem provides a linear relationship between regional production and specific factors of locations. In principle, the residual part of the estimation of Rybczynski, measured by the difference between the adjusted degree of explanation (R2) and the unit presents a approximated estimate of the importance not only of the "spillovers" effects, as considered by Kim (1999), but also of the industrial policies, because, industrial policies of modernization and innovation are interconnected with the "spillover" effects. However, it must be some caution with this interpretation, because, for example, although the growth of unexplained variation can be attributed to the growing importance of externalities "Marshallians" or "spillovers" effects and industrial policies, this conclusion may not be correct. Since the "spillovers" effects and industrial policies are measured as a residual part, the growth in the residual can be caused, also, for example, by growth in the randomness of the location of the products manufactured and the growing importance of external trade in goods and factors.

The designation of externalities "Marshallians" or "spillovers" effects, is because (2)Marshall (1920) have identified in a systematic way the causes of geographical concentration in industrial activities. Some recent works in these areas of thought, have sought to precisely identify solutions to these and other assumptions, present in some works, such as (3)Krugman (1991), (4)Krugman (1992), (5)Krugman et al. (1995), (6)Fujita et al. (1999), (7)Matsuyama (1991), (8)Gali (1994) and other related with spatial issues. It is also important to note that Marshall identifies the natural advantages as a major cause of geographical concentration (Kim, 1999). Another classification for the causes of concentration was made by (9)Scitovsky (1954), in other words, this author made a distinction between technological externalities, materialized through interactions unrelated to the market that directly affect consumers' utility or production functions of companies and pecuniary externalities that result from market interactions, in the proportion that companies engage in trade of products.

## 2. THE MODEL THAT ANALYZES THE IMPORTANCE OF NATURAL ADVANTAGES AND LOCAL RESOURCES IN AGGLOMERATION

According to Kim (1999), the Rybczynski theorem states that an increase in the supply of one factor leads to an increased production of the good that uses this factor intensively and a reduction in the production of other goods.

Given these assumptions, the linear relationship between regional output and offers of regional factors, may be the following:

$$Y = A^{-1}V,$$

where Y (nx1) is a vector of output, A (nxm) is a matrix of factor intensities or matrix input Rybczynski and V (mx1) is a vector of specific factors to locations.

For the output we used the gross value added of different manufacturing industries, to the specific factors of the locations used the labor, land and capital. For the labor we used the employees in manufacturing industries considered (symbolized in the following equation by "Labor") and the capital, because the lack of statistical data, it was considered until 1994, as a "proxy", the production in construction and public works (the choice of this variable is related to several reasons including the fact that it represents a part of the investment made during this period and symbolize the part of existing local resources, particularly in terms of infrastructure, because this has been considered from 1995) and from 1995 it was considered also production in this sector and the gross formation of fixed capital in manufacturing. With regard to land, although this factor is often used as specific of thr locations, the amount of land is unlikely to serve as a significant specific factor of thr locations. Alternatively, in this work is used the prodution of various extractive sectors, such as a "proxy" for the land. These sectors until 1994, include agriculture, forestry and fisheries (represented by "Agriculture") and production of natural resources and energy (symbolized by "Energy") and from 1995 include agriculture and fisheries (represented by "Agriculture") the forest ("Forest"), extractive industry of metallic mineral products ("Extraction1"), extractive industry of various products ("Extraction2") and energy production ("Energy"). The overall regression is then used as follows:

$$\ln Y_{it} = \alpha + \beta_1 \ln Labor_{it} + \beta_2 \ln Agriculture_{it} + \beta_3 \ln Florestry_{it} + \beta_4 \ln Extraction1_{it} +$$
$$\beta_5 \ln Extraction2_{it} + \beta_6 \ln Energy_{it} + \beta_7 \ln Construction_{it} + \beta_8 \ln Capital_{it} + \varepsilon$$

In this context, it is expected that there is, above all, a positive relationship between the production of each of the manufacturing industry located in a region and that region-specific factors required for this industry, in particular, to emphasize the more noticeable cases, between food industry and agriculture, among the textile industry and labor (given the characteristics of this industry), among the industry of metal products and metal and mineral extraction and from the paper industry and forest.

### 3. STATISTICAL DATA USED

Taking into account the variables of the model presented previously, we used statistical data of the five temporal regions of mainland Portugal, from the regional database of Eurostat statistics (Eurostat Regio of Statistics 2000) and INE (National Accounts 2003). The data are relative to regional gross value added of agriculture, fisheries and forestry, extractive industry of metallic mineral products, extractive industry of various products, the industry of fuel and energy products, each of manufacturing industry and construction and public works. We used also data relating to employees in each of the manufacturing industries and gross formation of fixed capital.

### 4. THE ESTIMATES MADE

In the estimates made were used panel data relating to the five regions of Portugal, for the years 1980-1999. It is intended to estimate the Rybczynski equation matrix in the linear form presented earlier. In the results presented in the following tables, there is a strong positive relationship between gross value added and labor in particular in the industries of metals, mineral products, chemicals, equipment and electrical goods, food, textile, paper and several products (which in some industries, it should not be checked). On the other hand, there is an increased dependence on natural and local resources in industries as the food and paper sectors. We found that the location of manufacturing industry is yet mostly explained by specific factors of locations, with a tendency to increase in some cases and poorly explained by "spillovers" effects and industrial policies. It should be noted, finally, little correlation between the location of manufacturing industry and gross formation of fixed capital in different regions of mainland Portugal.

**Table 1**: Results of estimations for the years 1980-1985

$$\ln Y_{it} = \alpha + \beta_1 \ln Labor_{it} + \beta_2 \ln Agriculture_{it} + \beta_3 \ln Energy_{it} + \beta_4 \ln Construction_{it} + \varepsilon$$

|  | IMT (2) | IMI (1) | IPQ (2) | IEE (1) | IET (1) | IAL (2) | ITE (2) | IPA (2) | IPD (1) |
|---|---|---|---|---|---|---|---|---|---|
| $\alpha$ | 4.975 (0.362) |  | 544.618 (1.144) |  |  | 1.219 (0.435) | -24.371(*) (-2.307) | 5.131 (0.644) | 245.405(*) (2.333) |
| **Dummy1** |  | 4.445 (0.916) |  | 10.717 (1.680) | 14.424 (0.815) |  |  |  |  |
| **Dummy2** |  | 5.482 (1.113) |  | 9.681 (1.544) | 15.662 (0.908) |  |  |  |  |
| **Dummy3** |  | 5.266 (1.056) |  | 10.858 (1.695) | 19.609 (1.109) |  |  |  |  |
| **Dummy4** |  | 5.064 (1.113) |  | 8.400 (1.395) | 9.376 (0.563) |  |  |  |  |
| **Dummy5** |  | 4.555 (1.078) |  | 7.052 (1.223) | 0.882 (0.054) |  |  |  |  |
| $\beta_1$ | 1.121(*) (4.685) | 0.147 (0.591) | -2.362(*) (-2.092) | 0.087 (0.317) | -2.630(*) (-3.533) | 0.742(*) (3.533) | 0.932(*) (5.388) | 1.180(*) (5.581) | 0.862(*) (3.089) |
| $\beta_2$ | 0.217 (0.390) | 0.130 (0.852) | 0.296 (0.675) | 0.217 (0.682) | -0.231 (-0.392) | 0.016 (0.103) | 1.704(*) (3.275) | 0.432(**) (1.791) | 0.349(*) (3.504) |
| $\beta_3$ | 0.012 (0.125) | 0.013 (0.299) | -0.084 (-0.520) | -0.022 (-0.336) | -0.305 (-0.336) | 0.202(*) (4.073) | -0.196 (-1.127) | -0.058 (-0.704) | -0.096(*) (-2.552) |
| $\beta_4$ | -0.064 (-0.128) | 0.475(*) (4.563) | -0.372 (-0.828) | 0.207 (1.022) | 1.846(*) (2.513) | 0.334(**) (1.880) | 0.173 (0.370) | -0.258 (-0.749) | 0.006 (0.054) |
| Sum of the elasticities | 1.286 | 0.765 | -2.522 | 0.489 | -1.320 | 1.294 | 2.613 | 1.296 | 1.121 |
| $R^2$ adjusted | 0.982 | 0.994 | 0.990 | 0.996 | 0.980 | 0.934 | 0.997 | 0.968 | 0.999 |
| Residual part | 0.018 | 0.006 | 0.010 | 0.004 | 0.020 | 0.066 | 0.003 | 0.032 | 0.001 |
| Durbin-Watson | 1.867 | 2.238 | 1.768 | 1.960 | 1.736 | 1.455 | 1.715 | 1.832 | 2.731 |
| Hausman test | 1.441(a) | 91.076(b)(*) | 6.934(a) | 25.407(b)(*) | 45.077(b)(*) | 2.426(a) | 1.905(a) | (c) | (c) |

**For each of the industries, the first values correspond to the coefficients of each of the variables and values in brackets represent t-statistic of each; (1) Estimation with variables "dummies"; (2) Estimation with random effects; (*) coefficient statistically significant at 5% (**) Coefficient statistically significant at 10%; IMT, metals industries; IMI, industrial mineral;, IPQ, the chemicals industries; IEE, equipment and electrical goods industries; EIT, transport equipment industry; ITB, food industry; ITE, textiles industries; IPA, paper industry; IPD, manufacturing of various products; (a) accepted the hypothesis of random effects; (b) reject the hypothesis of random effects; (c) Amount not statistically acceptable.**

**Table 2:** Results of estimations for the years 1986-1994

$$\ln Y_{it} = \alpha + \beta_1 \ln Labor_{it} + \beta_2 \ln Agriculture_{it} + \beta_3 \ln Energy_{it} + \beta_4 \ln Construction_{it} + \varepsilon$$

|  | IMT (2) | IMI (1) | IPQ (1) | IEE (1) | IET (1) | IAL (2) | ITE (1) | IPA (1) | IPD (2) |
|---|---|---|---|---|---|---|---|---|---|
| $\alpha$ | 10.010 (0.810) |  |  |  |  | 34.31(*) (3.356) |  |  | 83.250(*) (5.412) |
| **Dummy1** |  | 18.753(*) (5.442) | -13.467(*) (-3.134) | 14.333(*) (2.811) | 9.183 (1.603) |  | 15.175(*) (3.652) | 17.850(*) (3.162) |  |
| **Dummy2** |  | 19.334(*) (5.733) | -12.679(*) (-2.930) | 13.993(*) (2.802) | 10.084(**) (1.766) |  | 14.904(*) (3.597) | 17.532(*) (3.100) |  |
| **Dummy3** |  | 19.324(*) (5.634) | -13.134(*) (-3.108) | 14.314(*) (2.804) | 10.155(**) (1.797) |  | 14.640(*) (3.534) | 18.586(*) (3.313) |  |
| **Dummy4** |  | 18.619(*) (5.655) | -11.256(*) (-2.599) | 14.022(*) (2.857) | 9.384 (1.627) |  | 15.067(*) (3.647) | 15.001(*) (2.654) |  |
| **Dummy5** |  | 17.860(*) (5.629) | -11.060(*) (-2.682) | 12.629(*) (2.653) | 7.604 (1.377) |  | 13.206(*) (3.344) | 13.696(*) (2.574) |  |
| $\beta_1$ | 1.420(*) (4.965) | 0.517(*) (4.651) | 1.098(*) (8.056) | 0.817(*) (7.695) | 0.397(*) (2.455) | 0.378(*) (2.000) | 0.809(*) (5.962) | -0.071 (-0.230) | 0.862(*) (10.995) |
| $\beta_2$ | 0.844 (1.353) | -0.358(*) (-2.420) | 0.709(*) (2.628) | -0.085 (-0.480) | -0.314 (-0.955) | -0.026 (-0.130) | -0.484(**) (-1.952) | -0.171 (-0.505) | -0.148 (-0.780) |
| $\beta_3$ | 0.431 (1.468) | -0.242(*) (-3.422) | 0.120 (0.721) | -0.084 (-0.876) | 0.147 (0.844) | -0.067 (-0.706) | -0.229(**) (-1.738) | -0.165 (-0.904) | -0.524(*) (-5.289) |
| $\beta_4$ | -1.459(*) (-4.033) | 0.359(*) (2.629) | 0.260 (1.185) | 0.061 (0.318) | 0.433(*) (2.066) | 0.166 (0.853) | 0.529(*) (2.702) | 0.427 (1.596) | -0.085 (-0.461) |
| Sum of the elasticities | 1.236 | 0.276 | 2.187 | 0.709 | 0.663 | 0.451 | 0.625 | 0.020 | 0.105 |
| $R^2$ adjusted | 0.822 | 0.993 | 0.987 | 0.996 | 0.986 | 0.968 | 0.997 | 0.983 | 0.999 |
| Residual part | 0.178 | 0.007 | 0.013 | 0.004 | 0.014 | 0.032 | 0.003 | 0.017 | 0.001 |
| Durbin-Watson | 1.901 | 2.246 | 1.624 | 1.538 | 2.137 | 1.513 | 2.318 | 1.956 | 2.227 |
| Hausman test | (c) | 115.873[(b)(*)] | 26.702[(b)(*)] | 34.002[(b)(*)] | 9.710[(b)(*)] | (c) | 34.595[(b)(*)] | 26.591[(b)(*)] | 1.083[(a)] |

**Table 3**: Results of estimations for the whole period 1980-1994

$$\ln Y_{it} = \alpha + \beta_1 \ln Labor_{it} + \beta_2 \ln Agriculture_{it} + \beta_3 \ln Energy_{it} + \beta_4 \ln Construction_{it} + \varepsilon$$

| | IMT (2) | IMI (1) | IPQ (2) | IEE (1) | IET (1) | IAL (1) | ITE (1) | IPA (1) | IPD (1) |
|---|---|---|---|---|---|---|---|---|---|
| $\alpha$ | 7.771 (1.064) | | -9.035[(**)] (-1.834) | | | | | | |
| Dummy1 | | 13.865[(*)] (6.368) | | 14.020[(*)] (3.769) | 7.510 (1.188) | 16.733[(*)] (4.812) | -0.014 (-0.002) | 11.998[(*)] (2.875) | 23.631[(*)] (4.473) |
| Dummy2 | | 14.441[(*)] (6.759) | | 13.517[(*)] (3.727) | 8.675 (1.380) | 16.220[(*)] (4.798) | -0.026 (-0.003) | 11.874[(*)] (2.834) | 22.437[(*)] (4.378) |
| Dummy3 | | 14.390[(*)] (6.632) | | 14.235[(*)] (3.857) | 8.851 (1.413) | 16.648[(*)] (4.856) | -0.275 (-0.032) | 12.674[(*)] (3.055) | 23.537[(*)] (4.645) |
| Dummy4 | | 13.943[(*)] (6.833) | | 13.943[(*)] (4.002) | 7.545 (1.219) | 14.137[(*)] (4.498) | 0.350 (0.044) | 9.455[(*)] (2.259) | 20.806[(*)] (4.344) |
| Dummy5 | | 13.386[(*)] (6.747) | | 12.018[(*)] (3.485) | 4.952 (0.823) | 12.997[(*)] (4.207) | -1.264 (-0.165) | 8.874[(*)] (2.212) | 21.577[(*)] (4.443) |
| $\beta_1$ | 1.441[(*)] (8.880) | 0.555[(*)] (5.817) | 0.707[(*)] (5.555) | 0.742[(*)] (7.617) | 0.091 (0.468) | 0.360[(*)] (2.528) | 0.861[(*)] (4.161) | 0.178 (0.753) | 0.951[(*)] (13.642) |
| $\beta_2$ | 0.658 (1.608) | -0.197[(*)] (-2.122) | 0.781[(*)] (2.698) | -0.094 (-0.671) | -0.249 (-0.706) | 0.009 (0.077) | 0.219 (0.500) | 0.190 (0.784) | 0.136 (1.091) |
| $\beta_3$ | 0.011 (0.082) | -0.135[(*)] (-3.826) | 0.253[(*)] (2.512) | -0.072 (-1.276) | 0.073 (0.663) | -0.025 (-0.545) | -0.150 (-1.084) | 0.001 (0.009) | -0.236[(*)] (-4.583) |
| $\beta_4$ | -0.761[(*)] (-3.199) | 0.309[(*)] (3.615) | 0.066 (0.352) | 0.112 (0.867) | 0.646[(*)] (3.069) | 0.130 (1.253) | 0.434 (1.331) | 0.047 (0.325) | -0.166 (-1.538) |
| Dummy 1986 | 0.126 (0.637) | 0.143[(*)] (2.672) | -0.177 (-1.100) | 0.054 (0.658) | 0.154 (0.849) | -0.008 (-0.111) | 0.501[(*)] (2.225) | 0.632[(*)] (4.726) | 0.010 (0.131) |
| Sum of the elasticities | 1.349 | 0.532 | 1.807 | 0.688 | 0.561 | 0.474 | 1.364 | 0.416 | 0.685 |
| $R^2$ adjusted | 0.923 | 0.991 | 0.905 | 0.994 | 0.975 | 0.992 | 0.976 | 0.984 | 0.993 |
| Residual part | 0.077 | 0.009 | 0.095 | 0.006 | 0.025 | 0.008 | 0.024 | 0.016 | 0.007 |
| Durbin-Watson | 1.809 | 2.067 | 2.032 | 2.191 | 2.141 | 1.808 | 1.864 | 1.720 | 2.509 |
| Hausman test | (c) | 11690.230[(b)(*)] | (c) | 130.680[(b)(*)] | 405.296[(b)(*)] | 712.672[(b)(*)] | (c) | 22.553[(b)(*)] | 361.937[(b)(*)] |

**Table 4**: Results of estimations for the whole period 1995-1999

$$\ln Y_{it} = \alpha + \beta_1 \ln Labor_{it} + \beta_2 \ln Agriculture_{it} + \beta_3 \ln Florestry_{it} + \beta_4 \ln Extraction1_{it} +$$
$$\beta_5 \ln Extraction2_{it} + \beta_6 \ln Energy_{it} + \beta_7 \ln Construction_{it} + \beta_8 \ln Capital_{it} + \varepsilon$$

| | IMT (2) | IMI (2) | IPQ (2) | IEE (2) | IET (2) | IAL (1) | ITE (1) | IPA (1) | IPD (1) |
|---|---|---|---|---|---|---|---|---|---|
| $\alpha$ | 3.476 (0.365) | 3.151 (0.403) | -126.876 (-1.572) | 64.626[(*)] (4.362) | 17.203 (0.395) | | | | |
| Dummy1 | | | | | | | | | |
| Dummy2 | | | | | | | | | |
| Dummy3 | | | | | | | | | |
| Dummy4 | | | | | | -3.137 (-1.740) | -1.212 (-2.826) | 0.687 (0.663) | -0.497 (-0.590) |

| Dummy5 | | | | | | | | | |
|---|---|---|---|---|---|---|---|---|---|
| $\beta_1$ | 1.294(*) (7.664) | 1.251(*) (13.829) | 1.800 (1.339) | -0.073 (-0.321) | 0.684 (0.640) | 0.072 (0.332) | 0.747(*) (11.372) | 1.320(*) (2.887) | 0.585(**) (2.141) |
| $\beta_2$ | 0.136 (0.778) | -0.078 (-0.452) | 3.558(*) (2.929) | -1.334(*) (-4.651) | -0.482 (-0.703) | 0.795(*) (2.996) | 0.408(**) (3.914) | -0.638 (-1.666) | -0.114 (-0.411) |
| $\beta_3$ | -0.356 (-1.730) | -0.267 (-1.682) | 2.306 (1.209) | -1.242(*) (-3.769) | -0.639 (-0.521) | 0.822(**) (3.502) | 0.498(*) (6.317) | 0.376(*) (4.689) | 0.258(**) (2.227) |
| $\beta_4$ | -0.161(**) (-2.024) | -0.064 (-1.073) | 0.568 (0.911) | -0.175 (-1.475) | -0.147 (-0.423) | 0.180(**) (3.164) | 0.107(*) (5.271) | 0.036 (0.532) | -0.084 (-1.025) |
| $\beta_5$ | 0.606(*) (4.819) | 0.411(*) (3.386) | 2.198(*) (2.755) | -1.039(*) (-4.951) | 0.120 (0.180) | 0.011 (0.057) | -0.273(**) (-3.729) | -0.384 (-1.462) | 0.163 (0.509) |
| $\beta_6$ | -0.215 (-1.802) | -0.042 (-0.437) | -3.058(*) (-3.196) | 0.257 (1.338) | 0.404 (0.540) | -0.352 (-1.599) | -0.562(*) (-6.689) | -0.046 (-0.265) | -0.214 (-1.035) |
| $\beta_7$ | -0.237 (-1.247) | -0.182 (-1.371) | 0.330 (0.273) | 0.995(*) (3.153) | 0.134 (0.146) | -0.185 (-0.655) | 0.139 (1.560) | 0.553 (1.848) | 0.470 (1.265) |
| $\beta_8$ | -0.036 (-1.538) | 0.038(**) (2.043) | 0.407(*) (2.105) | 0.087(**) (2.351) | 0.101 (0.964) | 0.004 (0.143) | 0.072(*) (7.404) | -0.036 (-0.997) | -0.017 (-0.387) |
| Sum of the elasticities | 1.031 | 1.067 | 8.109 | -2.524 | 0.175 | 1.347 | 1.136 | 1.181 | 1.047 |
| R² adjusted | 0.999 | 0.999 | 0.999 | 0.999 | 0.999 | 0.999 | 0.999 | 0.999 | 0.999 |
| Residual part | 0.001 | 0.001 | 0.001 | 0.001 | 0.001 | 0.001 | 0.001 | 0.001 | 0.001 |
| Durbin-Watson | 2.343 | 2.282 | 1.988 | 2.221 | 2.191 | 2.021 | 3.074 | 2.747 | 2.400 |
| Hausman test | (c) | (c) | (c) | (c) | (c) | (c) | 16.063(b)(*) | 33381.450(b)(*) | 197.160(b)(*) |

**5. SOME CONCLUSIONS**

Analyzing the results of the estimations made it appears that there is a strong positive relationship between gross value added and labor, particularly in the industries of metals, mineral products, chemicals, equipment and electrical goods, food, textile, paper and several products, which in fact is proved by the statistical analysis. On the other hand, there is an increased dependence on natural and local resources in industries, as the food and paper firms.

About the existence of economies to scale or not, little can be concluded by the values obtained from the sum of elasticities. On the other hand, our entry into the EEC had little effect on the Portuguese manufacturing industry, what must be a matter for reflection.

In short, be noted that the location of the Portuguese manufacturing industry is still mostly explained by specific factors of locations, with a tendency to increase in some cases, capital has little influence on this location and the industrial policies of modernization and innovation are not relevant, especially those that have come from the European Union, what is more worrying.